\documentclass[12pt]{article}
\usepackage{amsfonts}
\usepackage{amsmath}
\usepackage{graphicx}
\usepackage{amsmath}
\usepackage{float}

\textwidth=6.5in \hoffset=-.75in \voffset=-1in \numberwithin{equation}{section}
\numberwithin{figure}{section}

\begin{document}

\begin{titlepage}
\bigskip 
\vspace{1cm}
\begin{center}
{\Large \bf {Kerr/CFT Correspondence in the Low Energy Limit of Heterotic String Theory}}\\
\end{center}
\vspace{2cm}
\begin{center}
A. M. Ghezelbash{ \footnote{ E-Mail: masoud.ghezelbash@usask.ca}}
\\
Department of Physics and Engineering Physics, \\ University of Saskatchewan, \\
Saskatoon, Saskatchewan S7N 5E2, Canada\\
\vspace{1cm}
\end{center}

\begin{abstract}
We investigate the recently proposed Kerr/CFT correspondence in the context
of heterotic string theory. The Kerr/CFT correspondence states that the
near-horizon states of an extremal four (or higher) dimensional black hole
could be identified with a certain chiral conformal field theory under the
conjecture that the central charges from the non-gravitational fields vanish.
The corresponding Virasoro algebra is generated by a class of diffeomorphisms
which preserves the appropriate boundary conditions on the near-horizon
geometry. To understand the chiral conformal field theory, we consider the
class of extremal Kerr-Sen black hole (that contains three non-gravitational
fields) as a class of solutions in the low energy limit (effective field
theory) of heterotic string theory. We derive the expression of the conserved
charges for the extremal Kerr-Sen solutions that contain dilaton, abelian gauge
filed and antisymmetric tensor filed. We confirm and extend the validity of the
conjecture (that the central charges from the non-gravitational fields vanish)
for theories including antisymmetric tensor fields. We combine the calculated
central charges with the expected form of the temperature using the Cardy
formula to obtain the entropy of the extremal black hole microscopically; in
agreement with the macroscopic Bekenstein-Hawking entropy of the extremal black
hole.
\end{abstract}
\end{titlepage}\onecolumn 
\bigskip 

\section{Introduction}

For a long time, black holes have been an interesting theoretical system to understand the nature of quantum gravity. Despite a lot of efforts to explain and reproduce the Bekenstein-hawking entropy, the theory of black hole entropy is not complete. 

Recently, in the context of proposed Kerr/CFT correspondence \cite{stro}, the microscopic entropy of four-dimensional extremal Kerr black hole is calculated by studying the dual chiral conformal field theory associated with the diffeomorphisms of near-horizon geometry of the Kerr black hole \cite{add3,add4,add5,add6,add1,add2}. These diffeomorphisms preserve an appropriate boundary condition at the infinity. One important feature of this correspondence is that it doesn't rely on supersymmetry and string theory unlike the well known AdS/CFT correspondence \cite{ads1,ads2,ads3,ads4,ads5,ads6}.

The Kerr/CFT correspondence has been used in \cite{LU} and \cite{cvetic} to find the entropy of dual CFT for 
four and higher dimensional Kerr black holes in AdS spacetimes and gauged supergravity as well as five-dimensional BMPV black holes in  \cite{ISO}. Moreover the correspondence has been used in string theory D1-D5-P and BMPV black holes in \cite{Aze} 
and in the five dimensional Kerr black hole in G\"{o}del universe \cite{G1}.
The continuous approach to the extremal Kerr black hole is essential in the proposed correspondence. For example, in the case of Reissner-Nordstrom black hole the approach to extremality is not continuous \cite{car}. The rotating bubbles, Kerr-Newman black holes in (A)dS spacetimes and rotating NS5 branes have been considered in \cite{Aze2}, \cite{Stro2} and \cite{Naka}.

In all these works, the central charge is computed only from the gravitational tensor field while contributions from other fields like scalar and vector fields are neglected. This led the authors of \cite{COM}, to the conjecture that the central charge of extremal black holes comes only from the gravitational field. In \cite{COM}, the authors verified the conjecture for a class of four and five dimensional theories that their actions contain gravity, scalar fields and a multiple of $U(1)$ vector fields as well as two topological terms (given in terms of vector fields and corresponding field strengths). Moreover, the conjecture was verified in \cite{Stro2} for the Kerr-Newmann-(A)dS black hole in the Einstein-Maxwell theory with cosmological constant. In \cite{Stro2}, the authors showed that there is no contribution to the central charge of dual CFT from $U(1)$ gauge field in the Einstein-Maxwell theory with cosmological constant.

In this article, inspired with the above mentioned works,  we consider the class of extremal Kerr-Sen black hole as a class of solutions in the low energy limit (effective field theory) of heterotic string theory and show that the antisymmetric tensor field (as well as the other non-gravitational fields) does not contribute to the central charge of the dual CFT. Hence we extend the validity domain of the conjecture (proposed in \cite{COM}) to include the non-gravitational antisymmetric tensor fields. The Kerr-Sen black hole is an exact solution to the four-dimensional effective field theory of heterotic string theory with gauge field, dilaton field and antisymmetric tensor filed \cite{sen}. The Kerr-Sen black hole has been studied in \cite{KS4} and \cite{KS3} in regard to its hidden symmetries, null geodesics, photon capture and its singularities. Moreover Kerr-Sen black hole has been used to study black hole lensing in the strong deflection limit \cite{KS1} and the massive complex scalar field in the Kerr-Sen geometry has been considered in \cite{KS2}.

We apply the Kerr/CFT correspondence to the extremal Kerr-Sen black hole and should stress that the Kerr-Sen solutions contain an antisymmetric tensor field as well as a dilaton and a vector field. 
As we mentioned before, the Kerr-Sen solutions are the exact solutions to the four-dimensional effective action of heterotic string theory. The effective action is obtained by compactifiying six of the ten dimensions of string theory and we have not included any massless fields arising from compactification in the theory. 
It seems Kerr-Sen black hole might be considered as a sub-class of black hole solutions in \cite{Stro2}, but we should notice that in all the solutions that have been considered in \cite{Stro2}, there are not absolutely any solutions with antisymmetric tensor fields. 
In four dimensions, the antisymmetric tensor field is equivalent to a scalar (axion) and we may expect that the results (coming from Kerr/CFT correspondence) should not be very different from the results presented in \cite{Stro2}. However, due to the non-trivial coupling of the antisymmetric tensor field to the Chern-Simons three form, 
in this article, we try explicitly to extend the validity of the conjecture (that the central charge of dual CFT to extremal black hole comes only from gravitational part and first proposed in \cite{COM}) for theories that contain antisymmetric tensor fields.

The outline of this paper is as follows. In section \ref{sec:5Dreview},
we first review briefly the Kerr-Sen black hole and its associated physical quantities. We find the near-horizon geometry of extremal black hole by
using special coordinate transformations as well as the near-horizon limits of the other non-gravitational fields. We notice a delicate divergence cancellation in the near-horizon limit of the three-form field strength due to the presence of the Chern-Simons terms. In section \ref{sec:sol},
we calculate the different contibutions to the central charge of CFT; from gravitational field, dilaton, gauge field and the antisymmetric tensor field. We find that there are no contributions to the central charge of the CFT from non-gravitational fields. Finally we find the microscopic entropy of extremal Kerr-Sen black hole in the dual chiral conformal field theory and compare the results with the macroscopic Bekenstein-Hawking entropy. We conclude in section \ref{sec:con} with a summary of our results.

\section{Extremal Kerr-Sen Black Hole}

\label{sec:5Dreview}

In this section, we give a brief review of the Kerr-Sen black hole and then study its near-horizon geometry.

The effective action of heterotic string theory in four dimensions is given by
\begin{equation}
S=-\int d^4x \sqrt{-det \, G}e^{-\Phi}(-R+\frac{1}{12}H^2-G^{\mu\nu}\partial _\mu \Phi\partial _\nu \Phi+\frac{1}{8}F^2)\label{action1}
\end{equation}
where $H^2=H_{\mu\nu\rho}H^{\mu\nu\rho}$ and $F^2=F_{\mu\nu}F^{\mu\nu}$. 
In (\ref{action1}), $G_{\mu\nu}$ and $\Phi$ are the metric and the dilaton field respectively, $F_{\mu\nu}=\partial _\mu A_\nu-\partial _\nu A_\mu$ is the field strength for the gauge field $A_\mu$ associated with a $U(1)$ subgroup of $E_8 \times E_8$ and
\begin{equation}
H_{\mu\nu\rho}=\partial _\mu B_{\nu \rho}+\partial _\nu B_{\rho \mu}+\partial _\rho B_{\mu \nu}-\frac{1}{4}(A_\mu F_{\nu\rho}+A_\nu F_{\rho\mu}+A_\rho F_{\mu\nu})\label{threeformHversusCS}
\end{equation}
where the last three terms are the gauge Chern-Simons terms.
In Einstein frame, the Kerr-Sen black hole is given by \cite{sen}
\begin{eqnarray}
ds^2&=&-\frac{r^2+a^2\cos^2(\theta)-2mr}{r^2+a^2\cos^2(\theta)+2mr\sinh^2(\alpha/2)}dt^2\nonumber\\
&+&\frac{r^2+a^2\cos^2(\theta)+2mr\sinh^2(\alpha/2)}{r^2+a^2-2mr}dr^2\nonumber\\
&+&({r^2+a^2\cos^2(\theta)+2mr\sinh^2(\alpha/2)})d\theta^2-
\frac{4mra\cosh^2(\alpha/2)\sin^2(\theta)}{{r^2+a^2\cos^2(\theta)+2mr\sinh^2(\alpha/2)}}dtd\phi\nonumber\\
&+&\{(r^2+a^2)(r^2+a^2\cos^2(\theta))+2mra^2\sin^2(\theta)+4mr(r^2+a^2)\sinh^2(\alpha/2)+4m^2r^2\sinh^4(\alpha/2)\}\nonumber\\
&\times& \frac{\sin^2(\theta)}{r^2+a^2\cos^2(\theta)+2mr\sinh^2(\alpha/2)}d\phi ^2.\label{BH}
\end{eqnarray}
The dilaton and gauge field components are
\begin{eqnarray}
\Phi&=&-\ln \frac{r^2+a^2\cos^2(\theta)+2mr\sinh^2(\alpha/2)}{r^2+a^2\cos^2(\theta)}\\
A_t&=&\frac{2mr\sinh(\alpha)}{r^2+a^2\cos^2(\theta)+2mr\sinh^2(\alpha/2)}\\
A_\phi&=&\frac{-2mra\sinh(\alpha)\sin^2(\theta)}{r^2+a^2\cos^2(\theta)+2mr\sinh^2(\alpha/2)}
\end{eqnarray}
and the only non-vanishing component of antisymmetric tensor field is
\begin{equation}
B_{t\phi}=\frac{2mra\sinh^2(\alpha/2)\sin^2(\theta)}{r^2+a^2\cos^2(\theta)+2mr\sinh^2(\alpha/2)}
.\end{equation}

The black hole solution (\ref{BH}) has mass $M=m\cosh^2(\alpha/2)$, charge $Q=\frac{m}{\sqrt{2}}\sinh\alpha$ and angular momentum $J=ma\cosh^2(\alpha/2)$.
We rewrite the metric as 
\begin{eqnarray}
ds^2&=&-(1-\frac{2M\tilde{r}}{\rho^2})d\tilde{t}^2+\rho^2(\frac{d\tilde{r}^2}{\Delta}+d\theta^2)\nonumber\\
&-&\frac{4M\tilde{r}a}{\rho^2}\sin^2\theta d\tilde{t}d\tilde{\phi}+\{\tilde {r}(\tilde {r}+\varrho)+a^2+\frac{2M\tilde {r} a^2\sin^2\theta}{\rho^2}\}\sin^2\theta d\tilde \phi^2\label{BH2}
\end{eqnarray}
where $\rho^2=\tilde r(\tilde r+\varrho)+a^2\cos^2\theta$ and $\Delta=\tilde r(\tilde r+\varrho)-2M\tilde r+a^2$.
The parameter $\varrho$ is related to $m=M-\frac{Q^2}{2M}$ and $\alpha$ in (\ref{BH}) by $\varrho=2m\sinh^2(\alpha/2)=Q^2/M$. The dilaton , gauge field and the antisymmetric tensor field are given by
\begin{eqnarray}
\Phi&=&-\ln \frac{\tilde{r}(\tilde{r}+\varrho)+a^2\cos^2\theta}{\tilde{r}^2+a^2\cos^2\theta}\label{dil}\\
A_{\tilde{t}}&=&\frac{2\sqrt{2}\tilde{r}Q}{\rho ^2}\label{A1}\\
A_{\tilde{\phi}}&=&\frac{-2\sqrt{2}\tilde{r}Qa\sin^2\theta}{\rho ^2}\label{A2}\\
B_{\tilde{t}\tilde{\phi}}&=&\frac{\tilde{r}\varrho a\sin^2\theta}{\rho ^2}.\label{Anti}
\end{eqnarray}
We notice that for special value of $\varrho=0$, the metric (\ref{BH2}) reduces to Kerr black hole. 

Moreover, we should note that the presence of antisymmetric tensor field $B_{\mu\nu}$ makes the action (\ref{action1}) quite different from the actions considered in \cite{COM,Stro2}. Although the Kerr-Sen black hole solution (\ref{BH2}) looks to be included in the class of general extremal black hole solutions (which considered in \cite{Stro2}), but we should mention that in all solutions considered in \cite{Stro2}, there are no antisymmetric tensor fields. The antisymmetric tensor field may or may not contribute to the central charge of the dual CFT and this is the main question that we try to address in this paper.

Moreover, we note that the metric (\ref{BH2}) is quite distinct from Kerr metric; it can not be obtained simply by a shift in coordinate $r$ from Kerr metric, hence we can not tell a priori about the outcome of applying Kerr/CFT correspondence to the rotating black hole solution (\ref{BH2}) with the dilaton, gauge field and especially the non-trivial antisymmetric tensor field (\ref{Anti}). On the other hand, for $a=0$, the Kerr-Sen black hole reduces to Gibbons-Maeda-Garfinkle-Horowitz-Strominger charged black hole of heterotic string theory in the strong deflection limit. In this limit, the Gibbons-Maeda-Garfinkle-Horowitz-Strominger charged black hole can be used to study the gravitational lensing when light passes close to the black hole.
The event horizon of black hole (\ref{BH2}) is
\begin{equation}
r_H=M-\frac{\varrho}{2}+\frac{1}{2}\sqrt{(2M-\varrho)^2-4a^2}.
\end{equation}
To avoid any naked singularity, we should impose
\begin{equation}
\mid J \mid \leq M^2 -\frac{1}{2}Q^2.
\end{equation}
It is obvious that the extremal black hole satisfies the upper bound of above inequality.
The angular velocity at the horizon and  Hawking temperature of black hole (\ref{BH}) are 
\begin{eqnarray} 
\Omega _H&=&\frac{a}{m(m+\sqrt{m^2-a^2})(1+\cosh(\alpha))}\\
T_H&=&\frac{\sqrt{m^2-a^2}}{2\pi m(m+\sqrt{m^2-a^2})(1+\cosh(\alpha))}
\end{eqnarray}
respectively. For the black hole in the form (\ref{BH2}), the corresponding angular velocity at horizon, Hawking temperature and entropy are 
\begin{eqnarray} 
\Omega _H&=&\frac{J}{M(2M^2-Q^2+\sqrt{(2M^2-Q^2)^2-4J^2})}\\
T_H&=&\frac{\sqrt{(2M^2-Q^2)^2-4J^2}}{4\pi M(2M^2-Q^2+\sqrt{(2M^2-Q^2)^2-4J^2})}\\
S&=&2\pi M(M-\frac{Q^2}{2M}+\sqrt{(M-\frac{Q^2}{2M})^2-\frac{J^2}{M^2}}).\label{entropy1}
\end{eqnarray}
We notice in the extremal limit where $J=M^2-\frac{1}{2}Q^2$, the angular velocity and Hawking temperature reduce to $\frac{1}{2M}$ and $0$, respectively and the entropy (\ref{entropy1}) reduces simply to $S=2\pi J$; independent of the mass of black hole.

To find the near-horizon limit of the extremal black hole, we change the coordinates according to the following transformations 
\begin{eqnarray}
\tilde r&=&(M-\frac{\varrho}{2})(1+\frac{\lambda}{y})\\
\tilde t&=& \frac{2M}{\lambda}t \\
\tilde \phi &=& \phi + t/\lambda
\end{eqnarray}
where the scaling parameter $\lambda$ approaches zero. The black hole metric (\ref{BH2}) changes to the near-horizon metric
\begin{eqnarray}
ds^2&=&\frac{(2M-\varrho)\{\frac{1}{2}\varrho\sin^2\theta+M(1+\cos^2\theta)\}^2}{2M(1+\cos^2\theta)+\varrho\sin^2\theta}
(\frac{-dt^2+dy^2}{y^2})\nonumber\\
&+&
\{M^2(1+\cos^2\theta)+\frac{1}{4}(-\varrho^2 \sin^2\theta-4\varrho M\cos^2\theta)\}d\theta^2\nonumber\\
&+&\frac{4(2M-\varrho)M^2\sin ^2\theta}{2M(1+\cos^2\theta)+\varrho\sin^2\theta}(d\phi+\frac{dt}{y})^2
\end{eqnarray}
or
\begin{eqnarray}
ds^2&=&\{M^2(1+\cos^2\theta)+\frac{1}{4}(-\varrho^2 \sin^2\theta-4\varrho M\cos^2\theta)\}
\{\frac{-dt^2+dy^2}{y^2}+d\theta^2+\nonumber\\
&+&\frac{4M^2\sin ^2\theta}{(\frac{1}{2}\varrho\sin^2\theta+M(1+\cos^2\theta))^2}(d\phi+\frac{dt}{y})^2\}
.\label{metricback}
\end{eqnarray}
The near-horizon metric definitely is not asymptotically flat. The near-horizon dilaton field is 
\begin{equation}
\Phi=\ln \frac{(2M^2-Q^2)(1+\cos^2\theta)}{Q^2\sin^2\theta+2M^2(1+\cos^2\theta)}
\label{dilatonback}
\end{equation}
and the near-horizon $U(1)$ field strength is given by
\begin{equation}
F=\frac{2\sqrt{2}Q(2M^2-Q^2)\sin^2\theta}{y^2(Q^2\sin^2\theta+2M^2(1+\cos^2\theta))}dy \wedge dt
-\frac{8\sqrt{2}M^2(2M^2-Q^2)\sin(2\theta)Q}{y(Q^2\sin^2\theta+2M^2(1+\cos^2\theta))^2}(yd\theta \wedge d\phi+d\theta \wedge dt).\label{Fonhor}
\end{equation}
Equation (\ref{Fonhor}) shows that near-horizon gauge field is 
\begin{equation}
A=-\frac{2\sqrt{2}Q(2M^2-Q^2)\sin^2\theta}{(Q^2\sin^2\theta+2M^2(1+\cos^2\theta))}(d\phi+\frac{dt}{y})
.\label{gaugeback}
\end{equation}
In the near-horizon limit, the three-form field strength $H_{\mu\nu\sigma}$ is 
\begin{equation}
H=\{{\cal H}\frac{dy}{y^2}-\frac{1}{y}{\cal H}'d\theta\}\wedge dt \wedge d\phi\label{threeformH}
\end{equation}
where 
\begin{equation}
{\cal H}(\theta)=\frac{2(2M^2-Q^2)^2Q^2\sin^4\theta}{\{Q^2\sin^2\theta+2M^2(1+\cos^2\theta)\}^2}\label{calH}
.\end{equation}
To get this resut in the limit of $\lambda \rightarrow 0$, the contribution of gauge Chern-Simons terms in the three-form field strength (\ref{threeformHversusCS}) is very crucial. In the near-horizon limit, both antisymmetric tensor field and Chern-Simons terms contributions to the three-form field strength (\ref{threeformHversusCS}) diverge. These two divergences exactly cancel each other, hence we obtain the finite result (\ref{threeformH}) for the three-form field strength near the horizon. We can introduce the new antisymmetric tensor field ${\cal B}$ by
\begin{equation}
{\cal B}=-\frac{{\cal H}(\theta)}{y} dt \wedge d\phi 
\label{gauge2back}
\end{equation}
such that
\begin{equation}
H=d{\cal B}. \label{HversuscalB}
\end{equation}
To cover the whole near-horizon geometry, we use the global coordinates
\begin{eqnarray}
y&=&\frac{1}{\cos(\tau)\sqrt{1+r^2}+r}\\
t&=&y\sin(\tau)\sqrt{1+r^2}\\
\phi&=&\varphi +\ln (\frac{\cos(\tau)+r\sin(\tau)}{1+\sin(\tau)\sqrt{1+r^2}})
\end{eqnarray}
so the global near-horizon gauge field and the metric are 
\begin{equation}
A=-\frac{2\sqrt{2}Q(2M^2-Q^2)\sin^2\theta}{(Q^2\sin^2\theta+2M^2(1+\cos^2\theta))}(d\varphi+r{d\tau})
\end{equation}
and
\begin{eqnarray}
ds^2&=&\{M^2(1+\cos^2\theta)+\frac{1}{4}(-\varrho^2 \sin^2\theta-4\varrho M\cos^2\theta)\}
\{-(1+r^2)d\tau^2+\frac{dr^2}{1+r^2}+d\theta^2+\nonumber\\
&+&\frac{4M^2\sin ^2\theta}{(\frac{1}{2}\varrho\sin^2\theta+M(1+\cos^2\theta))^2}(d\varphi+rd\tau)^2\}\label{global1}
.\end{eqnarray}
Moreover, the near-horizon three-form field strength components are 
\begin{eqnarray}
H_{\tau r\varphi}&=&{\cal H}\\
H_{\tau \theta \varphi}&=&\cos(\tau)\sqrt{1+r^2}{\cal H '}\\
H_{\tau r \theta}&=&-\frac{\sin(\tau)}{\sqrt{1+r^2}}{\cal H '}\\
H_{r \theta \varphi}&=&\frac{r \sin(\tau)}{\sqrt{1+r^2}}{\cal H '}
\end{eqnarray}
where ${\cal H}$ is given in (\ref{calH}). 
In the case of vanishing $\varrho$, the metric becomes the near-horizon geometry of the Kerr solution, as in \cite{stro,Bardeen1}. For a fixed $\theta$, the near-horizon geometry is a quotient of warped AdS$_3$ which the quotient arises from identification of $\varphi$ coordinate. The isometry group of the geometry is $SL(2,R)\times U(1)$, where $U(1)$ is generated by the Killing vector $-\partial _\varphi$ and $SL(2,R)$ is generated by three Killing vectors,
\begin{eqnarray}
J_1&=&2\sin\tau\frac{r}{\sqrt{1+r^2}}\partial_\tau-2\cos\tau\sqrt{1+r^2}\partial_r+\frac{2\sin\tau}{\sqrt{1+r^2}}\partial_
\varphi\\
J_2&=&-2\cos\tau\frac{r}{\sqrt{1+r^2}}\partial_\tau-2\sin\tau\sqrt{1+r^2}\partial_r-\frac{2\cos\tau}{\sqrt{1+r^2}}\partial_
\varphi\\
J_3&=&2\partial_\tau
\end{eqnarray}

\section{Microscopic Entropy in Dual CFT}
\label{sec:sol}

We recall that asymptotic symmetry group of a spacetime is the group of allowed symmetries that obey the boundary conditions. As a result, the definition of the charge associated with a symmetry depends both on the action as well as boundary conditions. Hence, to compute the charges associated with asymptotic symmetry group of Kerr-Sen solution, we should consider all possible contributions from all different fields in the action (\ref{action1}). 
Asymptotic symmetries of the action (\ref{action1}) include diffeomorphisms $\xi$ such that 
\begin{eqnarray}
\delta _\xi \Phi&=&{\cal L}_\xi \Phi \label{ddil}\\
\delta _\xi A_\mu&=&{\cal L}_\xi A_\mu\\
\delta _\xi g_{\mu\nu}&=&{\cal L}_\xi g_{\mu\nu}\\
\delta _\xi {\cal B}_{\mu\nu}&=&{\cal L}_\xi {\cal B}_{\mu\nu}\label{danti}
\end{eqnarray}
as well as the following gauge transformations $\Lambda$ and $\Psi$ for $A_\mu$ and ${\cal B}_{\mu\nu}$ respectively,
\begin{eqnarray}
\delta _\Lambda A_\mu&=&\partial _\mu \Lambda\label{gauget1}\\
\delta _\Psi {\cal B}_{\mu\nu}&=&\partial _\mu \Psi _\nu - \partial _\nu \Psi _\mu .\label{gauget2}
\end{eqnarray}
In equations (\ref{ddil})-(\ref{danti}), 
the Lie derivatives of dilaton, gauge field, metric and antisymmetric tensor field ${\cal B}$) are 
\begin{eqnarray}
{\cal L}_\xi \Phi&=&\xi _\mu \nabla ^\mu \Phi \\
{\cal L}_\xi A_\mu&=&\xi ^\nu F_{\mu\nu}+\nabla _\mu (A_\nu \xi ^\nu)\\
{\cal L}_\xi g_{\mu\nu}&=&\nabla_\mu \xi _\nu+\nabla_\nu \xi _\mu \\
{\cal L}_\xi {B}_{\mu\nu}&=&{B}_{\mu \rho}\partial _\nu\xi ^\rho+{B}_{\rho\nu}\partial _\mu\xi ^\rho+\xi ^\rho\partial _\rho {B}_{\mu \nu}.
\end{eqnarray}

Hence, there are four contributions to the associated charge of asymptotic symmetry group of Kerr-Sen solution. The contributions come from gravitational tensor, dilaton, $U(1)$ gauge field and antisymmetric tensor field ${\cal B}_{\mu\nu}$. So we have
\begin{equation}
Q_{\zeta, \Lambda, \Psi}=\frac{1}{8\pi}\int_{\partial \Sigma} (k_{\zeta}^{g} [h;g]+k_\zeta ^ {\Phi}[h,\phi;g,\Phi]+k_{\zeta,\Lambda}^{A}[h,a;g,A]+k_{\zeta,\Psi}^{{\cal B}}[h,b;g,{\cal B}]) \label{charge1}
\end{equation}
where $h,a,b$ and $\phi$ mean the infinitesimal variations of $g, A, {\cal B}$ and $\Phi$ fields, respectively, and  $\partial \Sigma $ is the boundary of a spatial slice. We should note, thanks to equation (\ref{HversuscalB}), there is no contribution to the charge (\ref{charge1}) from Chern-Simons terms. 
The gravitational and dilaton contribution two-forms  
$k_{\zeta}^{g} [h;g]$ and $k_\zeta^ {\Phi}[h,\phi;g,\Phi]$ are given by \cite{Y1,Y2,Y3}
\begin{eqnarray}
k_{\zeta}^{g} [h;g]&=&-\delta {\textbf Q}_\zeta^g+{\textbf Q}_{\delta \zeta}^g+i_\zeta{\bf \Theta}[h]-{\textbf E}_{{\cal L}}[{\cal L}_\zeta g,h]\\
k_\zeta^ {\Phi}[h,\phi;g,\Phi]&=&-i_\zeta {\bf \Theta} _\Phi
\label{fourks} 
\end{eqnarray}
where 
${\bf \Theta}_\Phi=*(\phi d \Phi)$, ${\bf \Theta}[h]=\star \{ (D^\beta h_{\alpha \beta}-g^{\mu\nu}D_\alpha h_{\mu\nu}) dx ^\alpha \}$ and 
\begin{equation}
{\textbf E}_{{\cal L}}[{\cal L}_\zeta g,h]=\star \{ \frac{1}{2} h_{\alpha \gamma}
(D^\gamma \zeta _\beta+D_\beta \zeta ^\alpha ) dx ^\alpha \wedge dx ^\beta \}
\end{equation}
and ${\textbf Q}_\zeta^g$ is the Koumar two-form 
\begin{equation}
{\textbf Q}_\zeta^g=\frac{1}{2} \star (D_\mu \xi _\nu-D_\nu \xi _\mu) dx^\mu \wedge dx^\nu .
\end{equation}

The last two terms in equation (\ref{charge1}) are contributions of one-form gauge field $A$ and two-form ${\cal B}$ field to the charge. In general for a $\hat p$-form $P$ with the associated $(\hat p+1)$-form field strength $R$, the contribution is given by \cite{Y3}
\begin{equation}
k_{\zeta , \Pi}^{P} [h,p;g,P]=-\delta {\textbf Q}_{\zeta,\Pi}^P+{\textbf Q}_{\delta \zeta,\delta \Pi}^{P} -i_\zeta{\bf \Theta^P}-{\textbf E}_{{\cal L}}^P[{\cal L}_\zeta P+d\Pi,p]
\end{equation}
where
\begin{eqnarray}
{\bf \Theta^P}&=&p \wedge \star R \\
{\textbf E}_{{\cal L}}^P[{\cal L}_\zeta P+d\Pi,p]&=&\star \{ \frac{1}{2(\hat p-1)!}p_{\mu\rho_1\cdots\rho_{\hat p-1}}(
{\cal L}_\zeta P+d\Pi)_\nu^{\rho_1 \cdots \rho_{\hat p-1}}
dx^\mu \wedge dx^\nu \} 
\end{eqnarray}
and the two-form ${\textbf Q}_{\zeta,\Pi}^P$ is 
\begin{equation}
{\textbf Q}_{\zeta,\Pi}^P=(i_\zeta P+\Pi)\wedge \star R .
\end{equation}
The explicit expressions for the contributions to the charge (\ref{charge1}) from gravity, dilaton and Maxwell field are given in the Appendix. 
We find the contribution of antisymmetric tensor field as
\begin{eqnarray}
k_{\zeta,\Psi}^{{\cal B}}[h,b;g,{\cal B}]&=&\frac{1}{12}\{\zeta ^\lambda(\epsilon_{\mu\nu\rho\beta} b_{\lambda\alpha}+\epsilon_{\mu\nu\rho\alpha}b_{\beta\lambda}+\epsilon_{\mu\nu\rho\lambda}b_{\alpha\beta})
H^{\mu\nu\rho}\}dx^\alpha \wedge dx^\beta\nonumber \\ &-& \frac{1}{6}\epsilon_{\mu\nu\rho\sigma}\{ \frac{1}{2}\xi^\lambda b_{\nu\lambda}H^{\mu\nu\rho}+(\frac{1}{2}\xi^\lambda {\cal B}_{\nu \lambda}+\Psi _\nu)(\delta H^{\mu\nu\rho}+\frac{1}{2}hH^{\mu\nu\rho})\}dx^\nu \wedge dx^\sigma \nonumber\\
&+&\frac{1}{8}\epsilon^{\mu\nu}_{\,\,\,\,\,\rho\sigma}b_{\mu}^{\alpha}({\cal L}_\zeta {\cal B}+d\Psi)_{\nu\alpha}
dx^\rho\wedge dx^\sigma .\end{eqnarray}
We choose the proper boundary condition for the near-horizon metric as the same as one in \cite{stro}. Moreover, we choose the boundary conditions for the $U(1)$ gauge field 
\begin{equation}
a_\mu \sim {\cal O}(r,1/r^2,1,1/r) \label{gaugeboundary}
\end{equation}
and for the dilaton as 
\begin{equation} 
\phi \sim {\cal O}(1)\label{dilatonboundary}
\end{equation}
where the coordinates are $(\tau,r,\theta,\varphi)$. For the antisymmetric tensor field, we choose
\begin{equation}
b_{\mu\nu} \sim {\cal O} 
\left( \begin{matrix}  
0 & 1/r^2 & 1/r   & 1 \\ 
  &   0   & 1/r^2 & 1/r \\
  &       &  0    & 1/r \\
   &      &       & 0
\end{matrix} 
\right)
\label{bboundary}
\end{equation}
to make sure that the conserved charges of the theory remain finite. We can show that the near-horizon metric has a class of commuting diffeomorphisms, labeled by $n=0,\pm 1,\pm 2, \cdots$
\begin{equation}
\zeta_n=-e^{-in\varphi}(\partial_\varphi+inr\partial_r).\label{zeta1}
\end{equation}
This diffeomorphism generates a Virasoro algebra without any central charge
\begin{equation}
[\zeta _m,\zeta _n]=-i(m-n)\zeta_{m+n}
.\end{equation}
Under the action of diffeomorphism (\ref{zeta1}), the gauge field gets a $\varphi$-component that is of the order of $1$ at infinity. This is in contrast to the boundary condition (\ref{gaugeboundary}). To restore the boundary condition (\ref{gaugeboundary}), we should perform a gauge transformation with the gauge function 
\begin{equation}
\Lambda_n(\theta,\varphi)=-\frac{2\sqrt{2}Q(2M^2-Q^2)\sin^2\theta}{(Q^2\sin^2\theta+2M^2(1+\cos^2\theta))}e^{-in\varphi}
\end{equation}
Hence, under a combination of diffeomorphism transformation and gauge transformation, the gauge field at infinity behaves exactly as it is expected by (\ref{gaugeboundary}). 
Moreover, the Lie derivative of the antisymmetric tensor field $\cal B$ at infinity, has the following components
\begin{eqnarray}
{\cal L}_\zeta{\cal B}_{\tau r}&=&\frac{1}{2}e^{-i(\pi/2+n\varphi)}n\sin(\tau){\cal H}^2(\theta)+{\cal O}({\frac{1}{r^2}})\\
{\cal L}_\zeta{\cal B}_{\tau \varphi}&=&\frac{1}{2}e^{-i(\pi/2+n\varphi)}n^2\sin(\tau){\cal H}^2(\theta)r+{\cal O}(\frac{1}{r})
\label{BADBs}
\end{eqnarray}
that are not in agreement with the boundary condition (\ref{bboundary}). The only other non-zero component of 
${\cal L}_\zeta{\cal B}$ at infinity, is 
\begin{equation}
{\cal L}_\zeta{\cal B}_{r \varphi}=\frac{e^{-i(\pi/2+n\varphi)}n\sin(\tau){\cal H}(\theta)}{1+\cos(\tau)}\frac{1}{r^2}+{\cal O}(\frac{1}{r^4})
\end{equation}
which behaves smoother than what is supposed to be. To find agreement with the boundary condition (\ref{bboundary}), we do a compensational transformation (\ref{gauget2}) with the following $\Psi_\mu$ function,
\begin{eqnarray}
(\Psi_r)_n&=&\frac{1}{2}ine^{-in\varphi}{\cal H}^2(\theta)\cos(\tau)\\
(\Psi_\varphi)_n&=&\frac{1}{2}n^2e^{-in\varphi}{\cal H}^2(\theta)r\cos(\tau).\label{specialpsi}
\end{eqnarray}
So, we find a combination of diffeomorphism transformation and (\ref{gauget2}) yields an antisymmetric tensor field that behaves at infinity, in agreement with the imposed boundary condition (\ref{bboundary}). 

The charge (\ref{charge1}) generates the symmetry $(\zeta, \Lambda, \Psi)_n$ and the algebra of the asymptotic
symmetric group is given by the Dirac bracket algebra of these charges
\begin{eqnarray}
\{Q_{\zeta, \Lambda, \Psi} , Q_{\tilde \zeta, \tilde \Lambda, \tilde  \Psi}\}_{D.B.}
&=&(\delta _{\tilde \zeta}+\delta _{\tilde \Lambda}+\delta_{\tilde \Psi})Q_{\zeta, \Lambda, \Psi}\nonumber\\
&=&\frac{1}{8\pi}\int_{\partial \Sigma} 
(k_{\zeta}^{g} [{\cal L}_{\tilde \zeta}g;g]+k_\zeta ^ {\Phi}[{\cal L}_{\tilde \zeta}g,{\cal L}_{\tilde \zeta}\Phi;g,\Phi]\nonumber\\
&+&k_{\zeta,\Lambda}^{A}[{\cal L}_{\tilde \zeta}g,{\cal L}_{\tilde \zeta}A+d\tilde \Lambda;g,A]+k_{\zeta,\Psi}^{{\cal B}}[{\cal L}_{\tilde \zeta}g,{\cal L}_{\tilde \zeta}{\cal B}+d\tilde \Psi;g,{\cal B}]). \nonumber \\
&&
\end{eqnarray}
Taking the background geometry $\hat g$ and fields $\hat \Phi$, $\hat A$ and $\hat {\cal B}$ by (\ref{metricback}), (\ref{dilatonback}), (\ref{gaugeback}) and (\ref{gauge2back}), we obtain 
\begin{eqnarray}
\{Q_{\zeta, \Lambda, \Psi} , Q_{\tilde \zeta, \tilde \Lambda, \tilde  \Psi}\}_{D.B.}&=&Q_{[(\zeta, \Lambda, \Psi),(\tilde \zeta,\tilde \Lambda, \tilde \Psi)]}+\frac{1}{8\pi}\int_{\partial\Sigma}(k_{\zeta}^{\hat g} [{\cal L}_{\tilde \zeta}\hat g;\hat g]+k_\zeta ^ {\Phi}[{\cal L}_{\tilde \zeta}\hat g,{\cal L}_{\tilde \zeta}\hat \Phi;\hat g,\hat \Phi]\nonumber\\
&+&k_{\zeta,\Lambda}^{A}[{\cal L}_{\tilde \zeta}\hat g,{\cal L}_{\tilde \zeta}\hat A+d\tilde \Lambda;\hat g,\hat A]+k_{\zeta,\Psi}^{{\cal B}}[{\cal L}_{\tilde \zeta}\hat g,{\cal L}_{\tilde \zeta}\hat {\cal B}+d\tilde \Psi;\hat g,\hat {\cal B}]).\label{VIR}
\end{eqnarray}
A straightforward and lengthy calculation shows that the algebra of the asymptotic symmetry group is a Viraso algebra generated by $(\zeta, \Lambda, \Psi)_n$ with the central charge 
\begin{equation}
c=c_g+c_\Phi+c_A+c_{\cal B}.
\end{equation}
The four contributions to the central charge are generated by the last four central terms in (\ref{VIR}), respectively. Moreover, we find that the chosen boundary conditions for the metric tensor (as the same boundary condition  in \cite{stro}), dilaton (given by (\ref{dilatonboundary})) , gauge field (given by (\ref{gaugeboundary})) and antisymmetric tensor field (given by (\ref{bboundary})), keep all the conserved charges as well as the central charges completely finite. Explicitely, we find that
\begin{eqnarray}
\int_{\partial\Sigma}k_{\zeta_m}^{\hat g} [{\cal L}_{\tilde \zeta_n}\hat g;\hat g]&=&8\pi J(m^3-m)\delta_{m,-n}\\
\int_{\partial\Sigma}k_{\zeta_m} ^ {\Phi}[{\cal L}_{\tilde \zeta_n}\hat g,{\cal L}_{\tilde \zeta_n}\hat \Phi;\hat g,\hat \Phi]&=&0\\
\int_{\partial\Sigma}k_{\zeta_m,\Lambda_m}^{A}[{\cal L}_{\tilde \zeta_n}\hat g,{\cal L}_{\tilde \zeta_n}\hat A+d\tilde \Lambda;\hat g,\hat A]&=&0\\
\int_{\partial\Sigma}k_{\zeta_m,\Psi_m}^{{\cal B}}[{\cal L}_{\tilde \zeta_n}\hat g,{\cal L}_{\tilde \zeta_n}\hat {\cal B}+d\tilde \Psi;\hat g,\hat {\cal B}]&=&0
\end{eqnarray}
which yield
\begin{eqnarray}
c_g&=&12J\\
c_\Phi&=&0\\
c_A&=&0\\
c_{\cal B}&=&0.
\end{eqnarray}
These results explicitly show that the non-gravitational fields (including the antisymmetric tensor field) do not contribute to the central charge of the dual CFT. 
Replacing the Dirac brackets by commutators yields a quantum Virasoro algebra with the central charge
\begin{equation}
c=12J
\end{equation}
for the dual chiral CFT corresponding to Kerr-Sen black hole (\ref{BH2}).
To find the entropy of dual chiral CFT, we need to find Frolov-Thorne temperature \cite{Fro}.
A straightforward calculation shows 
\begin{equation}
T_{FT}=\frac{1}{2\pi}.
\end{equation}
Finally, we obtain the microscopic entropy in dual chiral CFT by using the Cardy relation, as
\begin{equation}
S=\frac{\pi^2}{3}cT_{FT}=2\pi J.
\end{equation}
This microscopic result for the entropy is exactly the same as macroscopic entropy of black hole (\ref{entropy1}) in the extremal limit.

Although Kerr-Sen black hole in the limit of $a\rightarrow 0$ reduces to Gibbons-Maeda-Garfinkle-Horowitz-Strominger charged black hole, but Kerr/CFT correspondence fails for Gibbons-Maeda-Garfinkle-Horowitz-Strominger black hole. This is quite reasonable since in derivation of microscopic entropy, we implicitly assumed that the angular velocity of the horizon is not zero. 

\section{Concluding Remarks}

\label{sec:con}
In this paper, we considered the class of extremal Kerr-Sen black holes in the low energy limit of heterotic string theory. We found the near-horizon metric of the black hole, as well as the near-horizon limits of the other non-gravitational fields of the theory by taking the near-horizon procedure. We found that the contribution of the Chern-Simons terms to the three-form field strength of the theory is very crucial. In fact, the contribution of the antisymmetric tensor field to the three-form field strength in near-horizon limit, is divergent. Moreover, the contribution of the Chern-Simons terms to the three-form field strength in near-horizon limit, also is divergent. However, these two divergences cancel out exactly when we consider both contributions to the three-form field stength. We found an important result that states near the horizon (which has the topology of warped AdS$_3$), the three-form field strength depends explicitely on a new antisymmetric tensor field, and not to the Maxwell gauge filed. By choosing the proper boundary conditions for the gravitational field, dilaton, gauge field and the antisymmetric tensor field, we found the diffeomorphisms that generate Virasoro algebra without any central charge. The generator of diffeomorphisms which is a conserved charge, can be used to construct an algebra under Dirac brackets. This algebra is the same as diffeomorphism algebra but just with some extra central terms. These central terms, in general contribute to the the central charge of the Virasoro algebra. We showed that the only non-zero contribution to the central charge of the dual conformmal field theory comes from gravitational field. So, we extended the validity of conjecture (that the central charges from the non-gravitational fields vanish) to theories that include the antisymmetric tensor fields. The central charge together with Frolov-Thorne temperature enable us to find the microscopic entropy of the extremal Kerr-Sen black hole in dual chiral CFT. The microscopic entropy is exactly the same
as macroscopic Bekenstein-Hawking entropy of the extremal black hole. Our work provides further supportive evidence in favor of a Kerr/CFT correspondence in the low energy limit of heterotic string theory that contains three non-gravitational fields.

\bigskip
\vspace{0.5cm}
{\Large \bf Acknowledgments}

\vspace{0.5cm}
This work was supported by the Natural Sciences and Engineering Research
Council of Canada.

\bigskip\vspace{0.5cm}
{\Large \bf Appendix}

\vspace{0.5cm}
The contribution to the charge (\ref{charge1}) from the gravitational tensor is
\begin{eqnarray}
k_{\zeta}^{g} [h;g]&=&-\frac{1}{4}\epsilon_{\mu\nu\rho\sigma}\{\zeta^\sigma\nabla^\rho h-\zeta^\sigma\nabla_\lambda h^{\rho\lambda}+\zeta_\lambda\nabla^\sigma h^{\rho\lambda}+\frac{1}{2}h\nabla ^{\sigma}\zeta^\rho\nonumber\\
&-&h^{\sigma\lambda}\nabla_\lambda\zeta^\rho+\frac{1}{2}h^{\lambda\sigma}(\nabla ^
\rho\zeta_\lambda+\nabla_\lambda\zeta^\rho)\}dx^\mu \wedge dx^\nu .\label{Kgravity}
\end{eqnarray}
The Maxwell contribution is given by
\begin{eqnarray}
k_{\zeta,\Lambda}^{A}[h,a;g,A]&=&\frac{1}{8\pi}\epsilon_{\alpha\beta\mu\nu}\{(-\frac{1}{2}hF^{\mu\nu}+2F^{\mu\rho}h_\rho ^\nu-\delta F^{\mu\nu})(\zeta^\sigma A_\sigma+\Lambda)-F^{\mu\nu}\zeta^\sigma a_\sigma -2F^{\sigma\mu}\zeta^\nu a_\sigma\}dx^\alpha \wedge dx^\beta \nonumber \\ &-&\frac{1}{8}\epsilon_{\alpha\beta}^{\mu\nu}a_\mu({\cal L}_\zeta A_\nu+\partial _\nu \Lambda)dx^\alpha \wedge dx ^\beta . \label{KMaxwell}
\end{eqnarray}
We should note that the last two terms in (\ref{Kgravity}) as well as in (\ref{KMaxwell}) vanish for an exact Killing vector and an exact symmetry, respectively.
Finally, the dilaton contribution is
\begin{equation}
k_\zeta ^ {\Phi}[h,\phi;g,\Phi]=-\frac{1}{6}\phi\epsilon ^\nu _{\rho\sigma\lambda}\zeta^\rho\partial _\nu \Phi dx ^\sigma \wedge dx ^\lambda.
\end{equation}


\vspace{0.5cm}

\begin{thebibliography}{99}
\bibitem{stro} 
M. Guica, T. Hartman, W. Song and A. Strominger, [arXiv:0809.4266].

\bibitem{add3}
S.N. Solodukhin, \textit{Phys. Lett.} \textbf{B454} (1999) 213. 

\bibitem{add4} 
S. Carlip, \textit{Phys. Rev. Lett.} \textbf{82} (1999) 2828. 

\bibitem{add5}
G. Barnich and F. Brandt, \textit{Nucl. Phys.} \textbf{B633} (2002) 3. 

\bibitem{add6}
%
G. Barnich and G. Compere, \textit{J. Math. Phys.} \textbf{49} (2008) 042901. 

\bibitem{add1}
M. Park, \textit{Nucl. Phys.} \textbf{B634} (2002) 339.

\bibitem{add2}
G. Kang, J. Koga and M. Park, \textit{Phys. Rev.} \textbf{D70} (2004) 024005.

\bibitem{ads1}
E. Witten, \textit{Adv. Theor. Math. Phys.} \textbf {2} (1998) 253.

\bibitem{ads2}
J. Maldacena, \textit{Adv. Theor. Math. Phys.} \textbf {2} (1998) 231.

\bibitem{ads3}
K. Sfetsos and K. Skenderis, \textit{Nucl. Phys.} \textbf{B517} (1998) 179.

\bibitem{ads4}
D.Z. Freedman, S.D. Mathur, A. Matusis and L. Rastelli, \textit{Nucl. Phys.} \textbf{B546} (1999) 96.

\bibitem{ads5}
S.S. Gubser, I.R. Klebanov and A.M. Polyakov, \textit{Phys. Lett.} \textbf {B428} (1998) 105.

\bibitem{ads6}
A.M. Ghezelbash, K. Kaviani, S. Parvizi and A.H. Fatollahi, 
\textit{Phys. Lett.} \textbf{B435} (1998) 291. 

\bibitem{LU} 
H. L\"{u}, J. Mei and C.N. Pope, \textit{JHEP} \textbf{0904} (2009) 054.

\bibitem{cvetic}
D.D.K. Chow, M. Cveti\u{c}, H. L\"{u} and C.N. Pope, [arXiv:0812.2918].

\bibitem{ISO}  H. Isono, T.-S. Tai and W.-Y. Wen, [arXiv:0812.4440].


\bibitem{Aze} 
T. Azeyanagi, N. Ogawa and S. Terashima, [arXiv:0812.4883].

\bibitem{G1} 
J.-J. Peng and S.-Q. Wu, \textit{Phys. Lett.} \textbf{B673} (2009) 216.

\bibitem{car} 
S.M. Carroll, M.C. Johnson and L. Randall, [arXiv:0901.0931].

\bibitem{Aze2} 
T. Azeyanagi, N. Ogawa and S. Terashima, \textit{JHEP} \textbf{0904} (2009) 061.

%
\bibitem{Stro2}
T. Hartman, K. Murata, T. Nishioka and A. Strominger, \textit{JHEP} \textbf{0904} (2009) 019.

\bibitem{Naka}
Y. Nakayama, \textit{Phys. Lett.} \textbf{B673} (2009) 272.

\bibitem{COM}
G. Compere, K. Murata and T. Nishioka, \textit{JHEP} \textbf{0905} (2009) 077.

\bibitem{sen}
A. Sen, \textit{Phys. Rev. Lett.} \textbf{69} (1992) 1006.

\bibitem{KS4}
K. Hioki, U. Miyamoto, \textit{Phys. Rev.} \textbf{D78} (2008) 044007. 

\bibitem{KS3}
A. Burinskii and G. Magli, \textit{Annals Israel Phys. Soc.} \textbf{13} (1997) 296.  

\bibitem{KS1}
G.N. Gyulchev and S.S. Yazadjiev, \textit{Phys. Rev.} \textbf{D75} (2007) 023006. 

\bibitem{KS2}
S.Q. Wu and X. Cai, \textit{J. Math. Phys.} \textbf{44} (2003) 1084. 

\bibitem{Bardeen1} J.M. Bardeen and G.T. Horowitz, \textit{Phys. Rev.} \textbf{D60} 
(1999) 104030.

\bibitem{Y1}
G. Barnich and G. Compere, \textit{Phys. Rev. Lett} \textbf{95} (2005) 031302.

\bibitem{Y2}
M. Banados, G. Barnich, G. Compere and A. Gomberoff, \textit{Phys. Rev.} \textbf{D73} (2006) 044006.

\bibitem{Y3}
G. Compere, [arXiv:0708.3153].

\bibitem{Fro}
V.P. Frolov and K.S. Thorne, \textit{Phys. Rev.} \textbf{D39} (1989) 2125.





\end{thebibliography}
\end{document}